\documentclass[lettersize,journal]{IEEEtran}
\IEEEoverridecommandlockouts
\usepackage{algorithmic}
\usepackage{array}
\usepackage[caption=false,font=normalsize,labelfont=sf,textfont=sf]{subfig}
\usepackage{textcomp}
\usepackage{stfloats}
\usepackage{url}
\usepackage{verbatim}
\usepackage{graphicx}
\usepackage{cite}
\usepackage{amsmath,amssymb,amsfonts}
\usepackage{graphicx}
\usepackage{textcomp}
\usepackage{xcolor}
\usepackage{braket}
\usepackage{qcircuit}
\usepackage[justification=centering]{caption} 
\usepackage{tikz}
\usetikzlibrary{shapes.geometric, backgrounds}
\usetikzlibrary{arrows.meta} 
\usetikzlibrary{arrows.meta, bending}
\usepackage{graphicx} 
\def\BibTeX{{\rm B\kern-.05em{\sc i\kern-.025em b}\kern-.08em
    T\kern-.1667em\lower.7ex\hbox{E}\kern-.125emX}}
\def\BibTeX{{\rm B\kern-.05em{\sc i\kern-.025em b}\kern-.08em
    T\kern-.1667em\lower.7ex\hbox{E}\kern-.125emX}}
\usepackage{balance}
\begin{document}
\title{Discrete-Time Quantum Walks: A Quantum Advantage for Graph Representation}
\author{
    \IEEEauthorblockN{Boxuan Ai}
    
    \IEEEauthorblockA{\textit{China Unionpay, Shanghai,China}}\\
}

\maketitle

\begin{abstract}
This paper presents a novel methodology that transforms discrete-time quantum walks into a graph embedding technique, offering a fresh perspective on graph representation methods.Through mathematical manipulations, the approach of this paper adeptly maps intricate graph topologies into the Hilbert space, which significantly enhances the efficacy of graph analysis and paves the way for sophisticated quantum machine learning tasks. This development promises to revolutionize the intersection of quantum computing and  graph theory , charting new frontiers in the application of quantum algorithms to graph computing and network science.
\end{abstract}

\begin{IEEEkeywords}
Quantum Walk, Graph Representation
\end{IEEEkeywords}

\section{Introduction}
Graph representation,  also known as graph embedding, has emerged as a pivotal technique in the realm of data science, offering a potent means to distill complex structural information into low-dimensional vector spaces while preserving intrinsic topological properties. This technology holds particular promise in the analysis of financial transaction networks, where it plays a crucial role in uncovering hidden patterns and illicit activities. Specifically, graph representation techniques have shown remarkable efficacy in detecting money laundering syndicates and unraveling credit card fraud communities, facilitating the identification of suspicious entities and transactions amidst intricate networks of financial interactions. By encoding the intricate relationships between accounts, transactions, and entities into compact, yet informative, representations, these methods empower financial institutions to enhance their fraud detection capabilities and safeguard against economic crimes. Furthermore, the versatility of graph representation extends beyond the financial sector, finding applications in diverse domains such as intelligent transportation systems and the Internet of Things (IoT), where it enables efficient analysis of complex networks, aids in anomaly detection, and supports informed decision-making processes. As such, graph representation techniques stand as a testament to the transformative potential of advanced data analytics in addressing contemporary challenges across multiple disciplines.

Traditional graph representation and traversal techniques, notably classical random walks, have long been employed for various graph-related tasks. Classical random walks, by their stochastic nature, can effectively explore a graph, transitioning from node to node based on predefined probabilities. This mechanism has proven useful in tasks such as network analysis, where the patterns of visitation frequencies offer insights into the graph's structural properties.However, when confronted with the exigencies of large-scale, highly interconnected datasets, classical methods often encounter limitations. This is where the concept of quantum walks emerges as a promising avenue for innovation. By harnessing the principles of quantum mechanics, specifically superposition and interference, quantum walks offer a fundamentally different approach to understanding the intricate topologies of graphs.

Quantum walk, as a powerful extension of classical random walk into the realm of quantum mechanics, has garnered significant attention in recent years due to its remarkable potential in advancing algorithmic efficiency and solving complex problems in various domains. This quantum analogue leverages the principles of superposition and interference inherent to quantum mechanics, enabling the walker to traverse multiple paths simultaneously, thereby significantly enhancing its performance over its classical counterpart. Recent research  has underscored the versatility and efficacy of quantum walk algorithms in tackling a broad spectrum of graph-theoretic challenges.In the context of graph theory, quantum walk algorithms have demonstrated remarkable advantages over classical methods in tasks such as shortest path finding, graph connectivity analysis, and network centrality measures. By encoding the adjacency matrix of a graph into quantum walk operators, these algorithms harness the exponential speedup offered by quantum parallelism, allowing for the exploration of graph properties with unprecedented efficiency. For instance, quantum walks have been successfully employed to accelerate the search for marked nodes within graphs, achieving a search rate comparable to $O(\sqrt{N})$ in the number of nodes, far surpassing the linear scaling of classical algorithms\cite{szegedy}.Moreover, quantum walks have been instrumental in designing novel quantum algorithms for optimization problems, transforming them into energy minimization tasks that can be efficiently tackled by the algorithm's exploratory nature. The interference effects inherent to quantum walks further amplify the probability of reaching optimal solutions, underscoring their unique computational advantage. Additionally, experiments utilizing various physical platforms, including superconducting circuits, nuclear magnetic resonance, and photonic systems, have successfully demonstrated the feasibility and scalability of quantum walk implementations, paving the way for their integration into practical quantum information processing systems.Thus, the exploration of quantum walks as a tool for graph representation is not merely an academic curiosity but a necessity driven by the demands of contemporary data-intensive science. 

The adoption of quantum walks for graph representation becomes even more imperative when considering their potential to support subsequent quantum machine learning tasks. Quantum machine learning, an emerging field that seeks to harness the power of quantum computation for data analysis and pattern recognition, inherently deals with complex, high-dimensional datasets. Graphs, as versatile structures for representing intricate relationships between data points, play a pivotal role in this context. By utilizing quantum walks for graph representation, we can effectively encode the rich topological information of these graphs into quantum states, which can then be seamlessly processed by quantum machine learning algorithms. This integration not only preserves the structural integrity of the data but also leverages the quantum advantages of parallelism and interference to enhance the learning capabilities of the models. Therefore, the necessity of employing quantum walks for graph representation is underscored by their ability to bridge the gap between the quantum realm of data representation and the advanced analytics offered by quantum machine learning, ultimately paving the way for more efficient and insightful data-driven discoveries.

\section{Background}
\noindent \label{sec:background}
Quantum computing is an area of computing that draws upon the principles of quantum mechanics to perform computational tasks. It represents a departure from traditional, classical computing, which is based on binary digits (bits) that are either in a state of 0 or 1. Quantum computing, on the other hand, utilizes quantum bits, or qubits, which can exist in a superposition of states, enabling parallelism and potentially exponential speedup in certain computational problems.

A qubit is the fundamental unit of quantum information. Unlike classical bits, qubits can be in a superposition of two states, conventionally labeled as $|0\rangle$ and $|1\rangle$. The state of a qubit can be represented by a vector in a two-dimensional complex vector space, known as the Hilbert space. The general state of a qubit is expressed as a linear combination of the basis states:

\begin{equation}
|\psi\rangle = \alpha|0\rangle + \beta|1\rangle
\end{equation}

\noindent where $|\psi\rangle$ is the quantum state, $\alpha$ and $\beta$ are complex coefficients, and $|0\rangle$ and $|1\rangle$ are the orthonormal basis states. The complex coefficients satisfy the normalization condition $|\alpha|^2 + |\beta|^2 = 1$, ensuring that the state is properly normalized. The ability of a qubit to be in a superposition of states is a cornerstone of quantum computing's computational power.When a quantum system is measured, it undergoes a collapse into one of the basis states. The probability of measuring a qubit in the state $|0\rangle$ is $|\alpha|^2$, and the probability of measuring it in the state $|1\rangle$ is $|\beta|^2$. This is consistent with the Born rule from quantum mechanics and is a fundamental aspect of quantum computation.

While single-qubit operations are fundamental, the true power of quantum computing emerges with the use of multi-qubit systems. A multi-qubit system allows for the creation of entanglement and the execution of more complex quantum algorithms. The state space of a multi-qubit system is the tensor product of individual qubit state spaces, which means that the number of possible states grows exponentially with the number of qubits.

For a system of $n$ qubits, the state can be represented as a vector in a $2^n$-dimensional Hilbert space. The general state of an $n$-qubit system can be expressed as a linear combination of the product states:

\begin{equation}
|\psi\rangle = \sum_{i_1=0}^{1} \sum_{i_2=0}^{1} \cdots \sum_{i_n=0}^{1} \alpha_{i_1 i_2 \cdots i_n} |i_1 i_2 \cdots i_n\rangle
\end{equation}

where $|\psi\rangle$ is the quantum state, $\alpha_{i_1 i_2 \cdots i_n}$ are complex coefficients, and $|i_1 i_2 \cdots i_n\rangle$ represents a basis state with $i_1, i_2, \ldots, i_n \in \{0, 1\}$.

\section{Methodology}
\subsection{Graph Encoding}\label{AA}
Let $G = (V, E)$ be a graph, where $V = \{v_1, \ldots, v_N\}$ denotes the set of nodes with cardinality $N$, and $E$ represents the set of edges. Define $A \in \mathbb{R}^{N \times N}$ as the adjacency matrix of $G$, whose element in the $i$-th row and $j$-th column, denoted $A_{i,j}$, indicates the presence of a transaction relationship between bank card accounts $v_i$ and $v_j$, formalized as:  
\begin{equation}  
    A_{i,j} = \begin{cases}     
    1 & \text{if } (v_i, v_j) \in E \\    
    0 & \text{if } (v_i, v_j) \notin E    
    \end{cases}    
\end{equation}  
  
The degree of a node, quantifying the number of edges incident to it in the graph, is defined for node $v_j$ as $d_j = \sum_{i=1}^{N} A_{i,j}$, which corresponds to the sum of all elements in the $j$-th column of matrix $A$.  
  
A Markov process is a stochastic process endowed with the "memoryless" property, implying that the system's next state is contingent solely upon the current state, independent of preceding states. This attribute permits a concise representation of the graph structure through state transitions, as each node considers only those directly connected to it, disregarding others. The probability transition matrix, denoted $P \in \mathbb{C}^{N \times N}$ for graph $G$, characterizes the transition probabilities in a Markov process and is represented as:  
\begin{equation}  
    P_{i,j} = \frac{A_{i,j}}{d_j}    
\end{equation}  
Here, $P_{i,j}$ signifies the probability of transitioning from node $v_i$ to node $v_j$, mathematically expressed as $P_{i,j} = P(X_t = s_j | X_{t-1} = s_i)$, with the row sum of the probability transition matrix being unity: $\sum_{i=1}^{N} P_{i,j} = 1$.  
  
Quantum state encoding refers to the procedure of embedding classical data into quantum states. Any node $v_i$ within the transaction graph $G$ can be represented by a quantum state $\ket{i}$, an $N \times 1$ dimensional column vector with a 1 in the $i$-th position and 0s elsewhere, adhering to the one-hot encoding principle:  
\begin{equation}  
    \ket{1} = \begin{bmatrix} 1 \\ 0 \\ \vdots \\ 0 \end{bmatrix}, \ket{2} = \begin{bmatrix} 0 \\ 1 \\ \vdots \\ 0 \end{bmatrix}, \cdots, \ket{N} = \begin{bmatrix} 0 \\ 0 \\ \vdots \\ 1 \end{bmatrix}    
\end{equation}  
Given a graph $G$ with $N$ nodes, a minimum of $n \geq \log_2{N}$ quantum bits are requisite to represent the quantum state $\ket{i}$ of an individual node $v_i$.

\subsection{Quantum Walk}\label{AA}
Random walk refers to the irregular movement of a walker within a specific route or area, while random walk on graph structures involves particles randomly jumping between nodes along the edges of the graph. Random walk can capture the topological structure of networks, and thus it is widely applied in graph analysis tasks such as graph embedding and node ranking. 

Quantum walk, proposed by Aharonov\cite{ahar}, differs from classical random walk. The quantum superposition property endows the walking particle with the ability to "split" its presence on the graph, allowing it to move simultaneously in all directions according to probabilities, thereby exploring multiple paths in different directions in parallel. The probability of the walking particle being at different nodes is described by the amplitude of the wave function, which updates as the particle moves. Leveraging this characteristic, quantum walk transforms complex graph data into quantum states while capturing the topological features of the network. Compared to classical random walk, quantum walk exhibits an exponential speedup in diffusion rate due to its unique coherence property.

Quantum walk can be classified into two types: discrete-time and continuous-time. Discrete-time quantum walk further includes two specific approaches: coin quantum walk and scattering quantum walk. Coin quantum walk primarily operates on the nodes of a graph and requires a coin space with a dimension equal to the degree of the nodes, making it more suitable for regular graphs\cite{regular}. On the other hand, scattering quantum walk encodes all edges of the graph into quantum ground states, eliminating the need for a coin operator\cite{nocoin}.

Each edge in the graph $G$ corresponds to two different ground states, and for each node $v_i$, there exist two non-overlapping subspaces: the subspace $C_{v_i}$ spanned by the ground states starting from $v_i$, and the subspace $\Omega_{v_i}$ spanned by the ground states ending at $v_i$. Quantum walks can be seen as mappings from $C_{v_i}$ to $\Omega_{v_i}$. Firstly, the graph $G$ is transformed into the quantum state $\ket{\Phi}$ by superposing the standard basis states $\ket{i,j}$:
\begin{equation}
   \ket{\Phi} = \sum_{i=1}^N \sum_{j=1}^N \phi_{i,j} \ket{i,j}
\end{equation}
where $\Phi \in \mathbb{C}^{N \times N}$, and the element in the $i$-th row and $j$-th column is $\phi_{i,j} = \left(\frac{P_{i,j}}{N}\right)^{1/2}$, satisfying the normalization requirement of quantum superposition. This represents the superposition of all possible paths starting from $v_i$:
\begin{equation}
\ket{\phi_i} = \sum_{j=1}^N \sqrt{P_{i,j}} \ket{j} 
\end{equation}
Mathematically, $\ket{\phi_i}$ can be regarded as the transpose of the square root of the $i$-th row of the transition matrix $P$. Specifically, $\ket{\phi_i} = \left[\begin{matrix} \sqrt{P_{i,1}} & \sqrt{P_{i,2}} & \cdots & \sqrt{P_{i,N}} \end{matrix}\right]^T$. By taking the tensor product of the state $\ket{i}$ and the corresponding $\ket{\phi_i}$, we obtain the projected state $\ket{\psi_i}$ of the Markov chain:
\begin{equation}
\ket{\psi_i} = \ket{i} \otimes \left(\sum_{j=1}^N \sqrt{P_{i,j}} \ket{j}\right) = \sum_{j=1}^N \sqrt{P_{i,j}} \ket{i,j} \quad 
\end{equation}
In quantum mechanics, a set of orthogonal and normalized vectors $\{\ket{\psi_i}: i=1,\ldots,N\}$ can span an $N$-dimensional Hilbert space $\Psi$. Define the projection operator $\Pi$, which projects any vector in the Hilbert space $H$ onto the space $\Psi$:
\begin{equation}
    \Pi = \sum_{i=1}^N \ket{\psi_i}\bra{\psi_i} 
\end{equation}
  The Grover diffusion operator is the core of the Grover quantum search algorithm. It enhances the quantum state through reflection transformation, which can be regarded as a "rotation" operation of the qubits. Based on the projection operator $\Pi$, the Grover operator $G$ can be constructed as:
\begin{equation}
    G = 2\Pi - I
\end{equation}

Among all operators that satisfy unitarity and permutation symmetry, the Grover operator is the farthest from the identity operator, which intuitively can speed up the diffusion process.
The swap operator $S$ can be used to exchange the memory of two quantum registers, such that $S\ket{i,j} = \ket{j,i}$:
\begin{equation}
     S = \sum_{i=1}^N \sum_{j=1}^N\ket{i,j}\bra{j,i} 
\end{equation}
The mathematical representation of the quantum walk process is the unitary evolution operator $U_{QW} \in \mathbb{C}^{N \times N}$. Applying $U_{QW}$ to the initial state $\ket{\Phi}$ achieves the unitary evolution of the walk state. Combining equations (10) and (11), the expression for $U_{QW}$ is constructed as:
\begin{equation}
    U_{QW} = SG = S(2\Pi - I) 
\end{equation}
After $U_{QW}$ acts on each node, the incoming edge state is mapped to the outgoing edge state. The state obtained after one step of quantum walk evolution from $\ket{i,j}$ is:
\begin{equation}
    \ket{i,j} \xrightarrow{U_{QW}} \left(\frac{2}{d_j} - 1\right)\ket{j,i} + \frac{2}{d_j} \sum_{k \neq i, (j,k) \in E} \ket{j,k}
\end{equation}
Equation (15) represents that after one step of quantum walk(Fig.1), the walker has moved from $v_i$ to $v_j$. In the next step, starting from $v_j$, there is a probability of $\left(\frac{2}{d_j} - 1\right)^2$ to return to $v_i$, and a probability of $\left(\frac{2}{d_j}\right)^2$ to reach another node $v_k$ that is adjacent to $v_j$ but different from $v_i$.

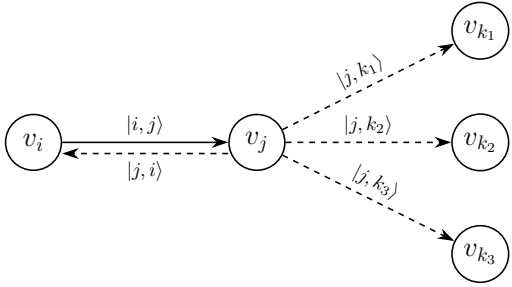
\begin{figure}[htbp]  
    \centering  
    \resizebox{0.38\textwidth}{!}{  
    \begin{tikzpicture}[auto, info/.style={font=\sffamily\Large\bfseries}]
    \node[circle, draw, thick, minimum size=1cm] [info] (v1) at (0,0) {$v_i$};
    \node[circle, draw, thick, minimum size=1cm, align=center] [info] (v2) at (4,0) {$v_j$};
    \node[circle, draw, thick, minimum size=1cm] [info] (v3) at (8,2) {$v_{k_1}$};
    \node[circle, draw, thick, minimum size=1cm, align=center] [info] (v4) at (8,0) {$v_{k_2}$};
    \node[circle, draw, thick, minimum size=1cm, align=center] [info] (v5) at (8,-2) {$v_{k_3}$};

    \draw[-{Stealth[length=3mm, width=2mm]}, thick] (v1) --node[above]{$|i,j\rangle$} (v2);
    \draw[-{Stealth[length=3mm, width=2mm]}, thick, dashed, transform canvas={yshift=-2mm}] (v2) --node[below]{$|j,i\rangle$} (v1);
    \draw[-{Stealth[length=3mm, width=2mm]}, thick, dashed] (v2) --node[above,sloped]{$|j,k_1\rangle$} (v3);
    \draw[-{Stealth[length=3mm, width=2mm]}, thick, dashed] (v2) --node[above]{$|j,k_2\rangle$} (v4);
    \draw[-{Stealth[length=3mm, width=2mm]}, thick, dashed] (v2) --node[above,sloped]{$|j,k_3\rangle$} (v5);
    \end{tikzpicture}
    }  
    \caption{Scattering Quantum Walk}  
\end{figure}
In a quantum computer, quantum walk does not require a real physical particle to move between nodes. Instead, it simulates the probability changes of a particle at different nodes by altering the states of quantum bits through quantum circuits and quantum gates. As the fundamental operation unit in quantum computation, quantum gates function similarly to logic gates in classical computation, both operating on basic units of information to change their states. Quantum walk utilizes quantum gates to implement unitary evolution on $n$ quantum bits, thereby adjusting the probabilities of the quantum bit superposition states and simulating the dynamic effect of a walking particle moving in a graph. The quantum walk at step $t$ is a unitary evolution operation applied to the current state $\ket{\Phi^{t-1}}$, expressed as follows:  
\begin{equation}
    U_{QW} \ket{\Phi^{t-1}} = \ket{\Phi^t}  
\end{equation}
When $t=0$, $\ket{\Phi^0}$ represents the initial quantum state of graph $G$, i.e., before the quantum walk begins. The preparation of the initial quantum state involves using quantum gates $U(\theta)$ to convert multiple low-energy state quantum bits $\ket{0}$ into a uniform superposition state as $\ket{\varphi_i^0}$, as shown in formula (2). The $U_{QW}$ quantum gate directly acts on the node $\ket{\varphi_i^{t-1}}$ as follows:  
\begin{equation}
    U_{QW} \ket{\varphi_i^{t-1}} = \ket{\varphi_i^t} 
\end{equation}  
The process of performing $t$ quantum walks on $\ket{\varphi_i^0}$ is as follows:  
\begin{equation}
    \ket{\varphi_i^0} \xrightarrow{U_{QW}} \ket{\varphi_i^1} \xrightarrow{U_{QW}} \cdots \ket{\varphi_i^{t-1}} \xrightarrow{U_{QW}} \ket{\varphi_i^t}  
\end{equation}
By repeatedly performing unitary evolution, the probability transition matrix $P$ also changes with each unitary evolution ($P^0 \rightarrow P^t$), allowing the quantum computer to simulate the process of a walking particle moving in a graph.  
The quantum circuit outputs the quantum states of all nodes sequentially, with the quantum states corresponding to nodes $v_1, \ldots, v_N$ represented as $\ket{\varphi_1^t}, \ket{\varphi_2^t}, \ldots, \ket{\varphi_N^t}$.  

\section{Discussion}
Compared to classical random walks, quantum walks offer several advantages in exploring graph topologies:  
  
\begin{enumerate}  
    \item \textbf{Quantum Superposition:} In classical random walks, the walker is limited to moving along a single path at any given time, resulting in significant information loss as topological details beyond the chosen path remain unexplored. Repeated walks to capture the full intricacies of the network can lead to substantial information redundancy. Conversely, quantum walks allow the walker to exist simultaneously at multiple positions and traverse multiple potential paths, enabling the exploration of the entire network within a minimal number of steps. This significantly enhances the efficiency and coverage of topological information mining.  
      
    \item \textbf{Quantum Interference:} Interference refers to the physical phenomenon where waves meet in space, resulting in superposition or cancellation and forming new waveforms. In the realm of quantum mechanics, the wave function describes the probability of a particle's existence at different positions. Quantum interference pertains to the interference between different wave functions. During a quantum walk, a particle can reach the same destination through multiple paths, each corresponding to a unique wave function. These wave functions superpose and interfere, enhancing or diminishing the probability of finding the particle at certain positions, thereby directly influencing the observation results. By leveraging the principle of quantum interference, quantum walks can sensitively detect minute details in graph structures, even amplifying slight differences within the graph.  
      
    \item \textbf{Complex Probability Amplitudes:} Classical random walks use real numbers to represent transition probabilities between nodes, while quantum walks employ complex probability amplitudes for movement. Complex probability amplitudes provide a richer representation of probability through both magnitude and phase, offering additional dimensions compared to real transition probabilities. The phase introduces an extra dimension that allows quantum walks to detect subtle structural characteristics of network topologies, such as local and global features. Therefore, quantum walks can extract significantly more information from graphs than classical random walks.  
\end{enumerate}  
In conclusion, this work presents a method for utilizing quantum walks in graph representation, offering a novel approach to encoding complex topological information into quantum states. By harnessing the unique properties of quantum mechanics, such as superposition and interference, my method provides a powerful tool for capturing the intricate relationships within graphs, paving the way for more efficient and insightful data analysis. This advancement lays the groundwork for subsequent work in quantum machine learning and related fields, where the ability to process high-dimensional, structured data is paramount. The integration of quantum walks into graph representation not only demonstrates the potential of quantum computation in enhancing traditional graph-based tasks but also opens up new avenues for exploring the complex web of connections that underlies much of the world's data. As such, this work stands as a significant step forward in the development of quantum-enhanced techniques for graph analysis and representation.

\end{document}